\documentclass{XrU2005}
\usepackage{graphicx}
\usepackage{natbib}

\newcommand{\gtsim}{\ {\raise-0.5ex\hbox{$\buildrel>\over\sim$}}\ }
\newcommand{\ltsim}{\ {\raise-0.5ex\hbox{$\buildrel<\over\sim$}}\ }

\makeatletter
\renewcommand{\section}{\@startsection%
{section}{1}{0mm}{-\baselineskip}%
{0.5\baselineskip}{\normalfont\Large\bfseries}}%
\makeatother

\title{Obscured and Unobscured AGN Evolution and the X-ray background}
\author[1,2]{Ezequiel Treister}
\author[1]{C. Megan Urry}
\author[2]{Paulina Lira}
\affil{Department of Astronomy, Yale University}
\affil[2]{Departamento de Astronom\'{\i}a, Universidad de Chile}

\begin{document}

%\keywords{diffuse radiation --- surveys: observations}

\maketitle

\begin{abstract}
\vspace{-0.6cm}
The Great Observatories Origins Deep Survey (GOODS) combines deep HST
and Spitzer imaging with the deepest Chandra/XMM observations to probe
obscured AGN at higher redshifts than previous multiwavelength
surveys. We present a self-consistent implementation of the AGN
unification paradigm, which postulates obscured AGN wherever there are
unobscured AGN, to successfully explain the infrared, optical, and
X-ray number counts of X-ray sources detected in the GOODS
fields. Assuming either a constant ratio of obscured to unobscured AGN
of 3:1 (the local value), or a ratio that decreases with luminosity,
and including Compton-thick sources, we can explain the spectral shape
and normalization of the extragalactic X-ray ``background'' as a
superposition of unresolved AGN, predominantly at $z$$\sim$0.5-1.5 and
L$_x$$\sim$10$^{43}$-10$^{44}$ ergs/s. The possible dependence of the
obscured to unobscured ratio with redshift is not well constrained;
present data allow it to decrease or increase substantially beyond
$z$$\sim$1.
\end{abstract}

\vspace{-0.6cm}
\section*{Method}
\vspace{-0.5cm}
The two main ingredients used to predict the AGN number counts and
contribution to the X-ray background are: ({\it i})The AGN luminosity
function and its evolution. We used the luminosity function of
\citet{ueda03}, which is based on hard X-ray observations and thus
relatively free of bias against obscured AGN. ({\it ii}) The AGN SED,
in terms of intrinsic luminosity and neutral hydrogen column density
($N_H$) along the line of sight. We assumed an underlying power-law
X-ray spectrum (E$>$0.5 keV) with photon index of $\Gamma$ 1.9,
typical of unobscured AGN. In the optical ($\lambda$=0.1-1 microns),
we used the Sloan Digital Sky Survey Composite Quasar Spectrum
\citep{vandenberk01} plus Milky-Way-type reddening laws and a standard
dust-to-gas ratio to convert $N_H$ to $A_V$. An L* elliptical host
galaxy was then added to the resulting optical spectrum. In the
infrared ($\lambda>$1 micron) we used dust emission models by
\citet{nenkova02} with the corresponding conversion from $N_H$ value
to viewing angle. AGN models with the same intrinsic luminosities were
normalized at 100 microns. The standard X-ray to optical luminosity
ratio was used to fix the scale of the different models.

\begin{figure}[ht!]
\includegraphics[width=0.45\textwidth, angle=0]{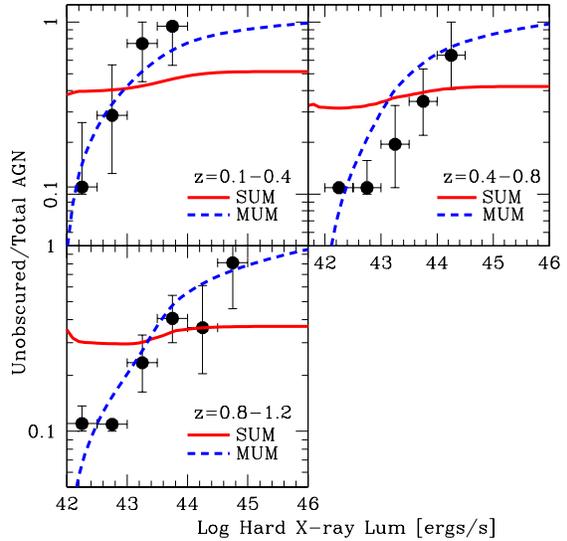}
\caption{\small Unobscured AGN ratio as a function of hard X-ray 
luminosity. {\it Filled circles}: data from \citet{barger05} combined with
our sample. ({\it Solid lines}:) Ratio predicted by our simple
unified model. The discrepancy between predictions and observations is
very clear. In this redshift range, the discrepancy cannot be
attributed to selection effects since the observed sample is mostly
complete both for obscured and unobscured AGN. Therefore, we also used
a modified unified model ({\it Dashed lines}) in which the
obscured-to-unobscured AGN ratio decreases linearly with
luminosity.}
\end{figure}

\vspace{-0.6cm}
\section*{Summary}
\vspace{-0.5cm}
Using the simplest AGN unification model we have explained the
spectral shape and intensity of the X-ray background. This is the
first demonstration that a model assuming a constant ratio of obscured
to unobscured AGN, independent of redshift or luminosity, can
simultaneously explain the observed X-ray background and the optical
and X-ray counts of AGN detected in deep X-ray surveys
\citep{treister04}. At the same time, a model that incorporates a
changing ratio with luminosity, as suggested by recently available
observations (Fig. 1), can also successfully explain the X-ray
background properties \citep{treister05}, as shown in Fig. 2.

\begin{figure}
\includegraphics[width=0.45\textwidth, angle=0]{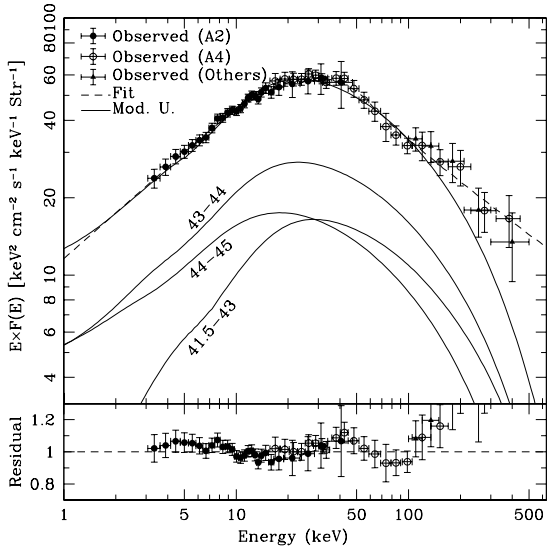}
\caption{X-ray background population synthesis for an AGN unification 
model in which the fraction of obscured AGN decreases with increasing
luminosity ({\it solid line}). The agreement with observations ({\it
data points, dashed line}) is very good, with a reduced $\xi^2$ of
0.648.  {\it Labeled solid lines} show the contribution from sources
in different X-ray luminosity bins. The maximum contribution to the
X-ray background comes from sources with log LX=43-44, that is,
moderate luminosity AGN.}
\end{figure}

The integral constraint of the X-ray background clearly does not
provide a sensitive probe of the fraction of the obscured AGN; other
observations, in particular a high-flux X-ray sample ($\gtsim
10^{-14}$~erg~cm$^{-2}$~s$^{-1}$) with a very high spectroscopic
completeness level ($>$90\%), is needed to test whether the ratio
depends on redshift, i.e., whether the evolution of obscured and
unobscured AGN is different.  As shown in Fig. 3, unification by
orientation --- in which obscured and unobscured AGN have the same
evolution --- is consistent with the data. In order to obtain this
highly complete sample, accurate redshifts for sources with optical
magnitudes $24\ltsim R\ltsim 27$ are needed. This is impossible with
current state-of-the-art 8m-class telescopes. However, good
photometric redshifts using medium-band filters down to these
magnitudes are possible and will allow to solve the redshift
dependence problem.

The resolved fraction of the X-ray background is $\ltsim 50\%$ in the
7-10 keV band and decreases with increasing energy. If the unification
model presented here is correct, $\sim50$\% of AGN are currently
missed by deep {\it Chandra} or {\it XMM} surveys. These are very
obscured AGN that will be detected only by hard X-ray observatories,
like the Black Hole Finder probe, at X-ray energies where the effects
of dust obscuration are negligible. These surveys will detect a large
fraction of the most obscured AGN, providing for the first time an
unbiased census of the black hole activity in the Universe.

ET would like to thank the support of Fundaci\'on Andes,
Centro de Astrof\'{\i}sica FONDAP and the Sigma-Xi
foundation through a Grant in-aid of Research. This work was
supported in part by NASA grant HST-GO-09425.13-A.

\begin{figure}
\includegraphics[width=0.45\textwidth, angle=0]{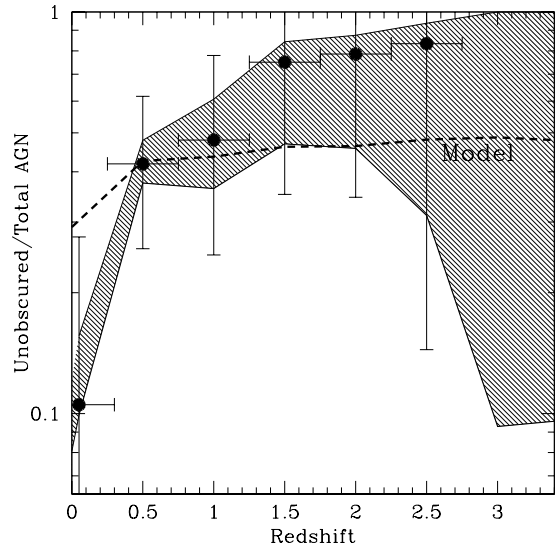}
\caption{Ratio of unobscured-to-total AGN as a
function of redshift. {\it Circles:} Observed data points from a
sample compiled from several X-ray surveys for which the total
spectroscopic completeness level is 66\%. {\it Thin solid lines:}
effects of adding the unidentified sources to each bin, weighting by
the comoving volume on each redshift bin, assuming that all the
unidentified sources are obscured (lower line) or unobscured (upper
line) AGN. {\it Dashed line}: Predicted ratio as a function of
redshift for a unified model in which the ratio (geometry) depends
only on luminosity, not redshift. That the data points lie
systematically above our model is not unexpected, since obscured AGN
are preferentially missed. The significant incompleteness at $z>1.5$
makes it impossible, given the current data, to rule out a constant
ratio with redshift.}
\end{figure}

\end{document}